\documentclass[aip,apl,reprint,showpacs]{revtex4-1}
\usepackage{graphicx}
\usepackage{changes}

\def\lSAW{\lambda_\mathrm{SAW}}

\begin{document}

\title{Acoustoelectric transport at gigahertz frequencies in coated epitaxial graphene}
\author{A. Hern\'andez-M\'inguez}
\email[e-mail address: ]{alberto.h.minguez@pdi-berlin.de}
\author{A. Tahraoui}
\author{J. M. J. Lopes}
\author{P. V. Santos}
\affiliation{Paul-Drude-Institut f\"ur Festk\"orperelektronik, Hausvogteiplatz 5-7, 10117 Berlin, Germany}

\date{\today}

\begin{abstract}
Epitaxial graphene (EG) produced from SiC surfaces by silicon sublimation is emerging as a material for electronic applications due to its good electronic properties and availability over large areas on a semiconducting substrate. In this contribution, we report on the transport of charge carriers in EG on SiC using high-frequency ($>$ 1 GHz) surface acoustic waves (SAWs). In our devices, the EG is coated with hydrogen-silsesquioxane, SiO$_2$ and a ZnO layer. This allows the efficient generation of SAWs and is compatible with the deposition of a metal top gate. Measurements of frequency- and time-resolved power scattering parameters confirm the generation and propagation of SAWs with frequencies of up to more than 7 GHz. Furthermore, the ZnO coating enhances the acoustoelectric currents by two orders of magnitude as compared to our previous uncoated samples. These results are an important step towards the dynamic acoustic control of charge carriers in graphene at gigahertz frequencies.
\end{abstract}

\maketitle

The strain and piezoelectric potentials accompanying surface acoustic waves (SAWs) provide a powerful tool for the dynamic modulation of the band structure in semiconductors, as well as for the transport and manipulation of elementary excitations in low-dimensional structures.\cite{Rocke97a, PVS152, PVS218, PVS246} In the case of a two-dimensional electron gas lying close to the surface of a piezoelectric substrate, the interaction between the electric charges and SAWs has been thoroughly studied,\cite{Parmenter_PR89_990_53, Hutson_JAP33_40_62, Ingebrigtsen_JAP41_454_70, Falko_PRB47_9910_93, Simon_PRB54_13878_96} and acoustoelectric currents have been experimentally observed.\cite{Esslinger_SSC84_939_92, Shilton_PRB51_14770_95,Shilton_JoPCM7_7675_95, Rotter_APL73_2128_98}
Recently, the interaction between SAWs and the pseudo-relativistic carriers in graphene has also been investigated.\cite{Thalmeier_PRB81_41409_10, Zhang_AA1_22146_11, Insepov_APL106_23505_15} Different experiments have demonstrated that SAWs can efficiently transport electric charges in a graphene layer transferred to the surface of an insulating piezoelectric substrate.\cite{Miseikis_APL100_133105_12, Bandhu_APL103_133101_13, Bandhu_APL105_263106_14, Roshchupkin_JAP118_104901_15} In addition, the use of dynamic SAW potentials for coupling light and plasmons has been proposed,\cite{Schiefele_PRL111_237405_13} as well as for the modulation of the graphene band structure.\cite{Dietel_PRB86_115450_12} These applications, however, require acoustic wavelengths comparable to the carrier mean free path, which is typically below a few hundred nanometers.

Among the various techniques nowadays available for graphene fabrication, epitaxial graphene (EG) on SiC is potentially advantageous for applications since it delivers large areas with reasonable mobilities. In addition, the material is directly on a semiconducting substrate, where it can be easily processed using conventional planar techniques. Recently, we have demonstrated that SAWs can induce unipolar electric currents in EG on SiC.\cite{Santos_APL102_221907_13} Due to the weak piezoelectricity of the SiC, however, the interaction between SAW and graphene carriers is relatively weak, yielding acoustoelectric currents of only a few pA. Future applications of SAWs in EG require efficient techniques for the generation of strong acoustic fields which, in addition, must be compatible with the control of the density and type of charge carriers.\cite{Bandhu_NR9_685_16}

\begin{figure}
\includegraphics[width=0.7\columnwidth]{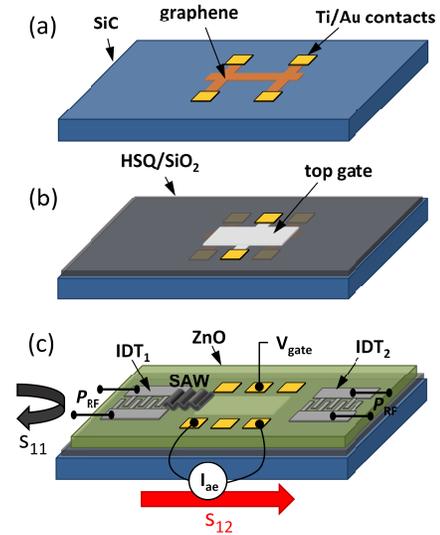}
\caption{Processing of SAW delay lines in coated graphene on SiC. After graphene etching and patterning of Hall bar structures (a), a dielectric layer of HSQ/SiO$_2$ is deposited prior to the evaporation of a semitransparent top metal gate (b). The samples are then coated with a piezoelectric ZnO layer and interdigital transducers (IDTs) for SAW excitation and detection are fabricated on the ZnO surface. Finally, the ZnO and HSQ/SiO$_2$ layers are selectively etched to acces the graphene contacts and gate (c). }\label{fig_setup}
\end{figure}

In this contribution, we study structures for the efficient generation of SAWs in EG with frequencies compatible with those of radio-frequency (RF) electronic devices ( $>$ 1 GHz). For that purpose, epitaxial monolayer graphene formed on the Si face of SiC substrates (10 mm $\times$ 10 mm size) by silicon sublimation\cite{Oliveira_APL99_111901_11} was etched into Hall bar structures of 10 $\mu$m width and 350 $\mu$m length, and contacted by Ti/Au metal pads for electric measurements, cf. Fig.~\ref{fig_setup}(a). The Hall bars were then spin-coated with a 100 nm-thick hydrogen-silsesquioxane (HSQ) layer and baked at 300 $^\circ$C. During the baking, the HSQ solidifies and its properties become similar to those of amorphous SiO$_2$. On top of it we deposited an additional 50 nm SiO$_2$ layer by sputtering. This HSQ/SiO$_2$ stack provides a suitable dielectric material for the application of vertical electric fields\cite{Hwang_JVSTB30_3_12} by a semitransparent top metal gate that allows for later optical characterization, cf. Fig.~\ref{fig_setup}(b). In addition, it protects the EG during the subsequent sputtering of a 350 nm-thick piezoelectric ZnO layer. Acoustic delay lines with single finger interdigital transducers (IDTs) deposited at the opposite edges of the graphene Hall bars are responsible for SAW excitation and detection, cf. Fig.~\ref{fig_setup}(c). The IDTs were designed for a fundamental acoustic wavelength $\lSAW=2.8~\mu$m, with 50~$\mu$m aperture. The center-to-center distance between IDT$_1$ and IDT$_2$ in Fig.~\ref{fig_setup}(c) defines the length of the delay line, $l_{delay}=1385$~$\mu$m. 

\begin{figure}
\includegraphics[width=0.7\columnwidth]{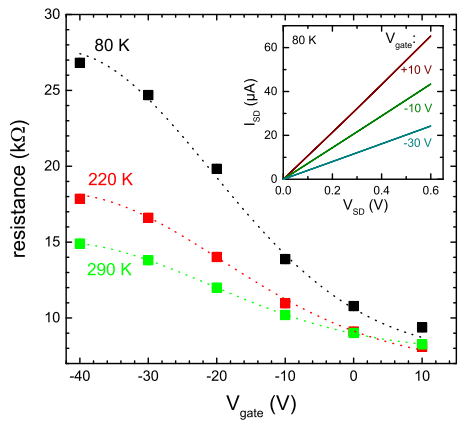}
\caption{Source-drain (SD) resistance as a function of gate bias, $V_\mathrm{gate}$, measured on a field-efect transistor in the structures of Fig.~\ref{fig_setup} at different temperatures. The dotted lines are a guide to the eye. The inset shows the source-drain current, $I_\mathrm{SD}$, as a function of the source-drain bias, $V_\mathrm{SD}$, for three different $V_\mathrm{gate}$ values, measured at 80 K.}\label{fig_charact}
\end{figure}

We have probed the impact of the fabrication process on the EG properties using both Raman spectroscopy\cite{APL_SM} and electric measurements. The electric properties are tested on a field-effect transistor with $14\times10~\mu$m$^2$ graphene area. Fig.~\ref{fig_charact} shows the source-drain (SD) resistance as a function of the voltage applied to the top gate, $V_\mathrm{gate}$, recorded at different temperatures. For that purpose, the sample was placed in vacuum in a low temperature probe station cooled by liquid nitrogen. The electrical resistance was obtained from the slope of the SD-current, $I_\mathrm{SD}$, as a function of the SD-voltage, $V_\mathrm{SD}$, at each $V_\mathrm{gate}$, cf. inset of Fig.~\ref{fig_charact}. As the measurements were performed in a two-contact configuration, they include the contribution of the contact resistance between the EG and the Ti/Au contacts, which is typically of several k$\Omega$. The resistance increases with negative bias, which indicates that the EG is $n$-doped. Taking into account that the gate voltage $V_0$ required to bring the Fermi level to the Dirac point is $V_0<-40$~V, we estimate that the electron concentration is $n_0=\vert V_0\vert C/e>4\times10^{12}$~cm$^{-2}$. Here, $e$ is the electron charge and $C=\epsilon_0\epsilon_r/d=177~\mu$Fm$^{-2}$ is the gate capacitance, obtained using a total thickness of the HSQ/SiO$_2$ layer of $d$=150~nm and a relative dielectric constant of $\epsilon_r\approx3$.\cite{Yuan_PSaT15_86_13} The value of $n_0$ is in accordance with the electron densities in our uncoated samples.\cite{Schumann_PRB85_235402_12}

The field-effect carrier mobility in the coated EG was estimated from Fig.~\ref{fig_charact} using the same procedure as in Ref.~\onlinecite{Bandhu_NR9_685_16}. It increases from 100 cm$^2$/Vs at room temperature to 160 cm$^2$/Vs at 78 K, and is therefore ten times lower than in uncoated EG at millikelvin temperature.\cite{Schumann_PRB85_235402_12} It is, however, one order of magnitude larger than in graphene mechanically transferred to a piezoelectric insulating substrate like LiNbO$_3$.\cite{Bandhu_APL105_263106_14,Bandhu_NR9_685_16}

\begin{figure}
\includegraphics[width=0.7\columnwidth]{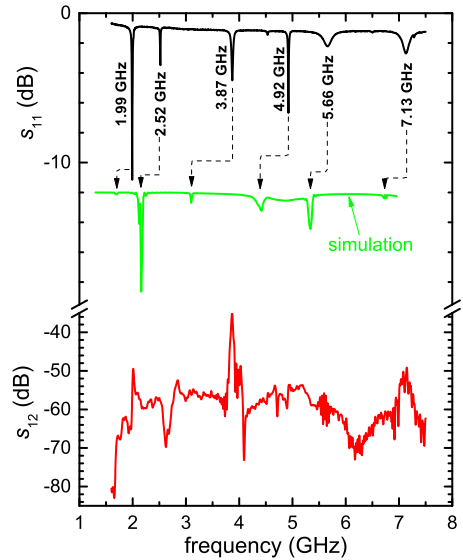}
\caption{Power reflection coefficient, $s_{11}$ (black curve), and transmission coefficient, $s_{12}$ (red curve), as a function of the applied RF frequency. The green curve displays the numerically simulated $s_{11}$ for IDTs with the same finger periodicity as the real sample. The simulation is vertically shifted for clarity.}\label{fig_SAWs_frequency}
\end{figure} 
  
We studied the SAW generation and propagation properties by measuring RF-power reflection ($s_{11}$) and transmission ($s_{12}$) coefficients of the IDTs at room temperature, cf. Fig.~\ref{fig_setup}(c). The $s_{11}$ spectrum (black curve) in Fig.~\ref{fig_SAWs_frequency} shows two fundamental acoustic modes with frequencies 1.99 and 2.52 GHz, as well as several higher frequency modes up to 7.13~GHz. Interestingly, the corresponding $s_{12}$ measurement (red curve) shows that only three of the modes generated by the IDT$_1$ (those with frequencies 1.99, 3.87 and 7.13 GHz) are efficiently detected by the IDT$_2$ after traversing the Hall bar. Several factors must be taken into account to understand this result. First, there is a frequency mismatch between IDT$_1$ and IDT$_2$ in the delay line, which reduces the sensitivity of the $s_{12}$ measurements. This is attributed to thickness variation of the HSQ, SiO$_2$ and ZnO layers along the sample surface. At higher frequencies, the reduction in sensitivity induced by the mismatch is partially compensated by a widening of the resonance peak. Second, we expect the SAW attenuation caused by scattering, by absorption at the different layers, and by the interaction of the SAW with the EG charge carriers, to be more intense for certain modes. A thorough study of this discrepancy is beyond the scope of this manuscript.

To understand the results of $s_{11}$, we have performed numerical simulations, cf. green curve in Fig.~\ref{fig_SAWs_frequency} (details of the calculations are described in the supplemental material\cite{APL_SM}). These reproduce the generation of the different modes, with two fundamental resonances at 1.70 and 2.18~GHz. We show in the supplemental material that the acoustic field of these modes is strongly confined in the overlayers due to the high contrast between their acoustic velocity and the one of the substrate.\cite{APL_SM} The latter accounts for the high electromechanical transduction as well as for the excitation of several overtones of the fundamental resonances.\cite{Nakahata_JJAP33_324_94, Nakahata_ITUFFC42_362_95, Naumenko_APL75_3029_99, Didenko_IToUFaFC47_179_, PVS120, Wu_JAP96_5249_04} In addition, the $\sim$64\% metallization ratio in our IDTs allows the generation of third harmonics of the fundamental wavelength\cite{Engan_ITED16_1014_69,Campbell_98} at frequencies 5.66 and 7.13~GHz (5.34 and 6.76~GHz in the simulation). The spatial profiles of these modes are also displayed in the supplemental material.\cite{APL_SM}

Although the simulations reproduce the generation of the different frequency modes, the measured resonances are blue-shifted with respect to the ones of the simulation. In addition, our calculations do not reproduce the amplitude of the resonances in the $s_{11}$  spectrum. In the case of the two fundamental modes, for example, their  relative amplitude is reversed with respect to the simulated ones, as well as to our previous uncoated devices, where no HSQ was used.\cite{Santos_APL102_221907_13} In our calculations, we use the nominal thickness of the deposited layers, and we treat the HSQ/SiO$_2$ stack as a single amorphous SiO$_2$ layer. Therefore the observed discrepancies between measured data and simulation could be related to experimental deviations from the nominal layer thickness. They could also be related to differences in the mechanical properties of the HSQ with respect to SiO$_2$ that are not fully accounted for in our model.

\begin{figure}
\includegraphics[width=0.7\columnwidth]{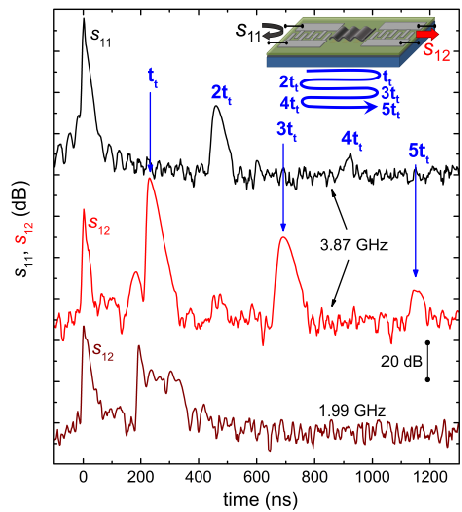}
\caption{Time-resolved profile of $s_{12}$ for the 1.99~GHz mode (dark red curve), as well as $s_{12}$ (red curve) and $s_{11}$ (black curve) for the 3.87~GHz mode. The vertical arrows mark the SAW transit time, $t_t\approx230$~ns, between IDT$_1$ and IDT$_2$. The curves are vertically shifted for clarity.}\label{fig_SAWs_time}
\end{figure} 

Time-resolved $s_{11}$ and $s_{12}$ spectra for modes with frequencies 1.99 and 3.87~GHz, acquired using a network analyzer with Fourier transform capability, are displayed in Fig.~\ref{fig_SAWs_time}. In the case of 1.99~GHz, only one echo appears at the transmission coefficient, $s_{12}$ (dark red curve), at a transit time $t_t\approx230$~ns. No echo is observed at $s_{11}$ (not displayed here). Taking into account that during $t_t$ the SAW travels a distance equal to $l_{delay}$, the acoustic group velocity can be estimated as $v_{group}\approx l_{delay}/t_t=6022$~m/s. The $s_{12}$ spectrum for 3.87~GHz (red curve) shows echoes not only at one transit time, but also at triple and quintuple transit times. These are attributed to multiple acoustic reflections at the IDTs, indicating that long living acoustic modes are efficiently generated in these structures. As expected, the corresponding echoes in $s_{11}$ (black curve) appear after two and four transit times. As shown in the supplemental material, up to 3 transit times are observed even for the 7.13~GHz mode.\cite{APL_SM}

\begin{figure}
\includegraphics[width=0.7\columnwidth]{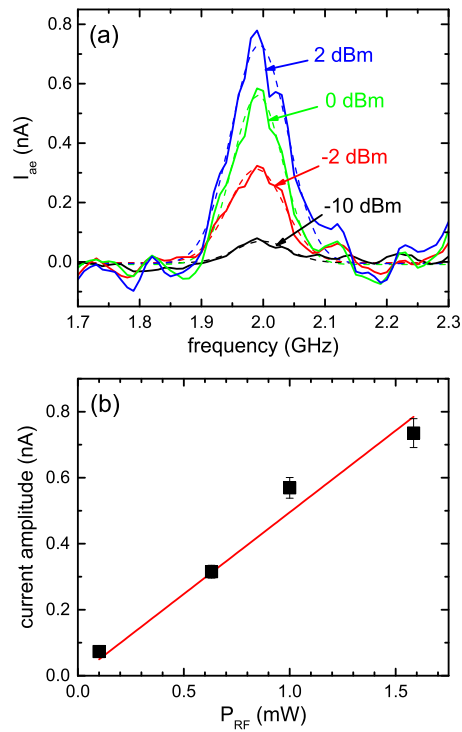}
\caption{(a) Spectral dependence of the acousto-electric current, $I_{ae}$ (cf. Fig.~\ref{fig_setup}(c)), induced in the graphene layer by the 1.99~GHz mode for several nominal RF-powers, $P_{RF}$, applied to the IDT. The dashed lines are Gaussian fits to the data. (b) Amplitude of the Gaussian fits of panel (a) as a function of $P_{RF}$. The red line is a linear fit to the data.}\label{fig_current}
\end{figure} 

We have measured the acoustoelectric current, $I_{ae}$, along the EG bar at room temperature for the different acoustic modes and compared it to the currents obtained previously in uncoated devices.\cite{Santos_APL102_221907_13} For each RF frequency and power applied continuously to the IDT, we measured $I_{ae}$ between two contacts separated 120 $\mu$m by using a Keithley ammeter. Of all modes, only the one at 1.99~GHz showed an acoustoelectric current above the sensitivity of our setup. Figure~\ref{fig_current}(a) displays the dependence of $I_{ae}$ on the frequency and nominal power, $P_{RF}$, of the RF signal applied to the IDT, while the squares in Fig.~\ref{fig_current}(b) show the amplitude of the Gaussian fit to the data vs. $P_{RF}$. As in our uncoated devices of Ref.~\onlinecite{Santos_APL102_221907_13}, the acoustoelectric current increases and is linearly proportional to $P_{RF}$ (red line in panel (b)). This is in agreement with the relaxation model for the interaction between SAWs and carriers in a two-dimensional electron gas.\cite{Bandhu_APL103_133101_13, Rotter_APL73_2128_98, Ingebrigtsen_JAP41_454_70, Falko_PRB47_9910_93} The acoustoelectric current in the ZnO-coated sample, however, reaches values of up to 0.8 nA. These current levels are about 300 times larger than the ones measured under similar SAW frequencies and nominal RF powers in uncoated samples, where the acoustic transport was attributed to the weak piezoelectric field induced by the SAW in the SiC substrate. We attribute this large current enhancement to the much stronger piezoelectric fields created by the top ZnO layer at the graphene region, which are estimated to be at least 20 times larger than the SiC contribution.

Finally, we have studied the effect of the top gate on the acoustoelectric current. As shown in the supplemental material, the free carriers in the metal gate locally screen the SAW piezoelectric field but not underneath.\cite{APL_SM} We did not observe, however, an effect of the gate voltage on $I_{ae}$. To modulate the acoustoelectric current by the top gate, the SAW-induced polarization charge density at the SiC/HSQ interface must be comparable to the graphene carrier density. For the maximum nominal RF-power used in our experiments, the estimated SAW-induced polarization charge density is, at least, 1000 times lower than the EG carrier density. In order to efficiently control $I_{ae}$ using the top gate, it is thus necessary to significantly reduce the carrier doping of the EG on SiC.

In summary, we have demonstrated the generation and propagation of SAWs on epitaxial monolayer graphene on SiC coated with a HSQ/SiO$_2$/ZnO film. Due to the acoustic velocity mismatch between the SiC and the top layers, high frequency modes up to 7 GHz are generated, including third harmonics with submicron wavelength. The stronger acoustic fields generated by the ZnO induce acoustoelectric currents substantially larger than the ones obtained in uncoated devices. This result is an important step towards the control of elementary excitations in graphene using dynamic potentials. Here, possible applications include the dynamic modification of the graphene conductivity by short range periodic modulations,\cite{Dietel_PRB86_115450_12} and the manipulation of the spin vector of the transported carriers by strain-induced gauge fields.\cite{Vozmediano_PR496_109_10}

\section*{Acknowledgments}
We thank J. M. Wofford for discussions and comments, and S. Rauwerdink for assistance in the preparation of the samples. This work was financially supported by the Deutsche Forschungsgemeinschaft (DFG) within the Priority Programme SPP 1459 Graphene.


\begin{thebibliography}{38}%
\makeatletter
\providecommand \@ifxundefined [1]{%
 \@ifx{#1\undefined}
}%
\providecommand \@ifnum [1]{%
 \ifnum #1\expandafter \@firstoftwo
 \else \expandafter \@secondoftwo
 \fi
}%
\providecommand \@ifx [1]{%
 \ifx #1\expandafter \@firstoftwo
 \else \expandafter \@secondoftwo
 \fi
}%
\providecommand \natexlab [1]{#1}%
\providecommand \enquote  [1]{``#1''}%
\providecommand \bibnamefont  [1]{#1}%
\providecommand \bibfnamefont [1]{#1}%
\providecommand \citenamefont [1]{#1}%
\providecommand \href@noop [0]{\@secondoftwo}%
\providecommand \href [0]{\begingroup \@sanitize@url \@href}%
\providecommand \@href[1]{\@@startlink{#1}\@@href}%
\providecommand \@@href[1]{\endgroup#1\@@endlink}%
\providecommand \@sanitize@url [0]{\catcode `\\12\catcode `\$12\catcode
  `\&12\catcode `\#12\catcode `\^12\catcode `\_12\catcode `\%12\relax}%
\providecommand \@@startlink[1]{}%
\providecommand \@@endlink[0]{}%
\providecommand \url  [0]{\begingroup\@sanitize@url \@url }%
\providecommand \@url [1]{\endgroup\@href {#1}{\urlprefix }}%
\providecommand \urlprefix  [0]{URL }%
\providecommand \Eprint [0]{\href }%
\providecommand \doibase [0]{http://dx.doi.org/}%
\providecommand \selectlanguage [0]{\@gobble}%
\providecommand \bibinfo  [0]{\@secondoftwo}%
\providecommand \bibfield  [0]{\@secondoftwo}%
\providecommand \translation [1]{[#1]}%
\providecommand \BibitemOpen [0]{}%
\providecommand \bibitemStop [0]{}%
\providecommand \bibitemNoStop [0]{.\EOS\space}%
\providecommand \EOS [0]{\spacefactor3000\relax}%
\providecommand \BibitemShut  [1]{\csname bibitem#1\endcsname}%
\let\auto@bib@innerbib\@empty
\bibitem [{\citenamefont {Rocke}\ \emph {et~al.}(1997)\citenamefont {Rocke},
  \citenamefont {Zimmermann}, \citenamefont {Wixforth}, \citenamefont
  {Kotthaus}, \citenamefont {B{\"o}hm},\ and\ \citenamefont
  {Weimann}}]{Rocke97a}%
  \BibitemOpen
  \bibfield  {author} {\bibinfo {author} {\bibfnamefont {C.}~\bibnamefont
  {Rocke}}, \bibinfo {author} {\bibfnamefont {S.}~\bibnamefont {Zimmermann}},
  \bibinfo {author} {\bibfnamefont {A.}~\bibnamefont {Wixforth}}, \bibinfo
  {author} {\bibfnamefont {J.~P.}\ \bibnamefont {Kotthaus}}, \bibinfo {author}
  {\bibfnamefont {G.}~\bibnamefont {B{\"o}hm}}, \ and\ \bibinfo {author}
  {\bibfnamefont {G.}~\bibnamefont {Weimann}},\ }\href@noop {} {\bibfield
  {journal} {\bibinfo  {journal} {Phys. Rev. Lett.}\ }\textbf {\bibinfo
  {volume} {78}},\ \bibinfo {pages} {4099} (\bibinfo {year}
  {1997})}\BibitemShut {NoStop}%
\bibitem [{\citenamefont {Stotz}\ \emph {et~al.}(2005)\citenamefont {Stotz},
  \citenamefont {Hey}, \citenamefont {Santos},\ and\ \citenamefont
  {Ploog}}]{PVS152}%
  \BibitemOpen
  \bibfield  {author} {\bibinfo {author} {\bibfnamefont {J.~A.~H.}\
  \bibnamefont {Stotz}}, \bibinfo {author} {\bibfnamefont {R.}~\bibnamefont
  {Hey}}, \bibinfo {author} {\bibfnamefont {P.~V.}\ \bibnamefont {Santos}}, \
  and\ \bibinfo {author} {\bibfnamefont {K.~H.}\ \bibnamefont {Ploog}},\
  }\href@noop {} {\bibfield  {journal} {\bibinfo  {journal} {Nat. Mater.}\
  }\textbf {\bibinfo {volume} {4}},\ \bibinfo {pages} {585} (\bibinfo {year}
  {2005})}\BibitemShut {NoStop}%
\bibitem [{\citenamefont {Couto{, Jr.}}\ \emph {et~al.}(2009)\citenamefont
  {Couto{, Jr.}}, \citenamefont {Lazi{\'c}}, \citenamefont {Iikawa},
  \citenamefont {Stotz}, \citenamefont {Hey},\ and\ \citenamefont
  {Santos}}]{PVS218}%
  \BibitemOpen
  \bibfield  {author} {\bibinfo {author} {\bibfnamefont {O.~D.~D.}\
  \bibnamefont {Couto{, Jr.}}}, \bibinfo {author} {\bibfnamefont
  {S.}~\bibnamefont {Lazi{\'c}}}, \bibinfo {author} {\bibfnamefont
  {F.}~\bibnamefont {Iikawa}}, \bibinfo {author} {\bibfnamefont
  {J.}~\bibnamefont {Stotz}}, \bibinfo {author} {\bibfnamefont
  {R.}~\bibnamefont {Hey}}, \ and\ \bibinfo {author} {\bibfnamefont {P.~V.}\
  \bibnamefont {Santos}},\ }\href@noop {} {\bibfield  {journal} {\bibinfo
  {journal} {Nat. Phot.}\ }\textbf {\bibinfo {volume} {3}},\ \bibinfo {pages}
  {645} (\bibinfo {year} {2009})}\BibitemShut {NoStop}%
\bibitem [{\citenamefont {Hern{\'a}ndez-M{\'i}nguez}\ \emph
  {et~al.}(2012)\citenamefont {Hern{\'a}ndez-M{\'i}nguez}, \citenamefont
  {M{\"o}ller}, \citenamefont {Breuer}, \citenamefont {Pf{\"u}ller},
  \citenamefont {Somaschini}, \citenamefont {Lazi{\'c}}, \citenamefont
  {Brandt}, \citenamefont {Garc{\'i}a-Crist{\'o}bal}, \citenamefont {de~Lima},
  \citenamefont {Cantarero}, \citenamefont {Geelhaar}, \citenamefont
  {Riechert},\ and\ \citenamefont {Santos}}]{PVS246}%
  \BibitemOpen
  \bibfield  {author} {\bibinfo {author} {\bibfnamefont {A.}~\bibnamefont
  {Hern{\'a}ndez-M{\'i}nguez}}, \bibinfo {author} {\bibfnamefont
  {M.}~\bibnamefont {M{\"o}ller}}, \bibinfo {author} {\bibfnamefont
  {S.}~\bibnamefont {Breuer}}, \bibinfo {author} {\bibfnamefont
  {C.}~\bibnamefont {Pf{\"u}ller}}, \bibinfo {author} {\bibfnamefont
  {C.}~\bibnamefont {Somaschini}}, \bibinfo {author} {\bibfnamefont
  {S.}~\bibnamefont {Lazi{\'c}}}, \bibinfo {author} {\bibfnamefont
  {O.}~\bibnamefont {Brandt}}, \bibinfo {author} {\bibfnamefont
  {A.}~\bibnamefont {Garc{\'i}a-Crist{\'o}bal}}, \bibinfo {author}
  {\bibfnamefont {M.~M.}\ \bibnamefont {de~Lima}}, \bibinfo {author}
  {\bibfnamefont {A.}~\bibnamefont {Cantarero}}, \bibinfo {author}
  {\bibfnamefont {L.}~\bibnamefont {Geelhaar}}, \bibinfo {author}
  {\bibfnamefont {H.}~\bibnamefont {Riechert}}, \ and\ \bibinfo {author}
  {\bibfnamefont {P.~V.}\ \bibnamefont {Santos}},\ }\href {\doibase
  10.1021/nl203461m} {\bibfield  {journal} {\bibinfo  {journal} {Nano Lett.}\
  }\textbf {\bibinfo {volume} {12}},\ \bibinfo {pages} {252} (\bibinfo {year}
  {2012})},\ \Eprint
  {http://arxiv.org/abs/http://pubs.acs.org/doi/pdf/10.1021/nl203461m}
  {http://pubs.acs.org/doi/pdf/10.1021/nl203461m} \BibitemShut {NoStop}%
\bibitem [{\citenamefont {Parmenter}(1953)}]{Parmenter_PR89_990_53}%
  \BibitemOpen
  \bibfield  {author} {\bibinfo {author} {\bibfnamefont {R.~H.}\ \bibnamefont
  {Parmenter}},\ }\href {\doibase 10.1103/PhysRev.89.990} {\bibfield  {journal}
  {\bibinfo  {journal} {Phys. Rev.}\ }\textbf {\bibinfo {volume} {89}},\
  \bibinfo {pages} {990} (\bibinfo {year} {1953})}\BibitemShut {NoStop}%
\bibitem [{\citenamefont {Hutson}\ and\ \citenamefont
  {White}(1962)}]{Hutson_JAP33_40_62}%
  \BibitemOpen
  \bibfield  {author} {\bibinfo {author} {\bibfnamefont {A.~R.}\ \bibnamefont
  {Hutson}}\ and\ \bibinfo {author} {\bibfnamefont {D.~L.}\ \bibnamefont
  {White}},\ }\href {\doibase 10.1063/1.1728525} {\bibfield  {journal}
  {\bibinfo  {journal} {J. Appl. Phys.}\ }\textbf {\bibinfo {volume} {33}},\
  \bibinfo {pages} {40} (\bibinfo {year} {1962})}\BibitemShut {NoStop}%
\bibitem [{\citenamefont {Ingebrigtsen}(1970)}]{Ingebrigtsen_JAP41_454_70}%
  \BibitemOpen
  \bibfield  {author} {\bibinfo {author} {\bibfnamefont {K.~A.}\ \bibnamefont
  {Ingebrigtsen}},\ }\href {\doibase http://dx.doi.org/10.1063/1.1658696}
  {\bibfield  {journal} {\bibinfo  {journal} {J. Appl. Phys.}\ }\textbf
  {\bibinfo {volume} {41}},\ \bibinfo {pages} {454} (\bibinfo {year}
  {1970})}\BibitemShut {NoStop}%
\bibitem [{\citenamefont {Fal'ko}, \citenamefont {Meshkov},\ and\ \citenamefont
  {Iordanskii}(1993)}]{Falko_PRB47_9910_93}%
  \BibitemOpen
  \bibfield  {author} {\bibinfo {author} {\bibfnamefont {V.~I.}\ \bibnamefont
  {Fal'ko}}, \bibinfo {author} {\bibfnamefont {S.~V.}\ \bibnamefont {Meshkov}},
  \ and\ \bibinfo {author} {\bibfnamefont {S.~V.}\ \bibnamefont {Iordanskii}},\
  }\href@noop {} {\bibfield  {journal} {\bibinfo  {journal} {Phys. Rev. B}\
  }\textbf {\bibinfo {volume} {47}},\ \bibinfo {pages} {9910} (\bibinfo {year}
  {1993})}\BibitemShut {NoStop}%
\bibitem [{\citenamefont {Simon}(1996)}]{Simon_PRB54_13878_96}%
  \BibitemOpen
  \bibfield  {author} {\bibinfo {author} {\bibfnamefont {S.~H.}\ \bibnamefont
  {Simon}},\ }\href {\doibase 10.1103/PhysRevB.54.13878} {\bibfield  {journal}
  {\bibinfo  {journal} {Phys. Rev. B}\ }\textbf {\bibinfo {volume} {54}},\
  \bibinfo {pages} {13878} (\bibinfo {year} {1996})}\BibitemShut {NoStop}%
\bibitem [{\citenamefont {Esslinger}\ \emph {et~al.}(1992)\citenamefont
  {Esslinger}, \citenamefont {Wixforth}, \citenamefont {Winkler},\ and\
  \citenamefont {Kotthaus}}]{Esslinger_SSC84_939_92}%
  \BibitemOpen
  \bibfield  {author} {\bibinfo {author} {\bibfnamefont {A.}~\bibnamefont
  {Esslinger}}, \bibinfo {author} {\bibfnamefont {A.}~\bibnamefont {Wixforth}},
  \bibinfo {author} {\bibfnamefont {R.~W.}\ \bibnamefont {Winkler}}, \ and\
  \bibinfo {author} {\bibfnamefont {J.~P.}\ \bibnamefont {Kotthaus}},\ }\href
  {\doibase 10.1016/0038-1098(92)90797-D} {\bibfield  {journal} {\bibinfo
  {journal} {Sol. State Comm.}\ }\textbf {\bibinfo {volume} {84}},\ \bibinfo
  {pages} {939} (\bibinfo {year} {1992})}\BibitemShut {NoStop}%
\bibitem [{\citenamefont {Shilton}\ \emph
  {et~al.}(1995{\natexlab{a}})\citenamefont {Shilton}, \citenamefont {Mace},
  \citenamefont {Talyanskii}, \citenamefont {Pepper}, \citenamefont {Simmons},
  \citenamefont {Churchill},\ and\ \citenamefont
  {Ritchie}}]{Shilton_PRB51_14770_95}%
  \BibitemOpen
  \bibfield  {author} {\bibinfo {author} {\bibfnamefont {J.~M.}\ \bibnamefont
  {Shilton}}, \bibinfo {author} {\bibfnamefont {D.~R.}\ \bibnamefont {Mace}},
  \bibinfo {author} {\bibfnamefont {V.~I.}\ \bibnamefont {Talyanskii}},
  \bibinfo {author} {\bibfnamefont {M.}~\bibnamefont {Pepper}}, \bibinfo
  {author} {\bibfnamefont {M.~Y.}\ \bibnamefont {Simmons}}, \bibinfo {author}
  {\bibfnamefont {A.~C.}\ \bibnamefont {Churchill}}, \ and\ \bibinfo {author}
  {\bibfnamefont {D.~A.}\ \bibnamefont {Ritchie}},\ }\href {\doibase
  10.1103/PhysRevB.51.14770} {\bibfield  {journal} {\bibinfo  {journal} {Phys.
  Rev. B}\ }\textbf {\bibinfo {volume} {51}},\ \bibinfo {pages} {14770}
  (\bibinfo {year} {1995}{\natexlab{a}})}\BibitemShut {NoStop}%
\bibitem [{\citenamefont {Shilton}\ \emph
  {et~al.}(1995{\natexlab{b}})\citenamefont {Shilton}, \citenamefont {Mace},
  \citenamefont {Talyanskii}, \citenamefont {Simmons}, \citenamefont {Pepper},
  \citenamefont {Churchill},\ and\ \citenamefont
  {Ritchie}}]{Shilton_JoPCM7_7675_95}%
  \BibitemOpen
  \bibfield  {author} {\bibinfo {author} {\bibfnamefont {J.~M.}\ \bibnamefont
  {Shilton}}, \bibinfo {author} {\bibfnamefont {D.~R.}\ \bibnamefont {Mace}},
  \bibinfo {author} {\bibfnamefont {V.~I.}\ \bibnamefont {Talyanskii}},
  \bibinfo {author} {\bibfnamefont {M.~Y.}\ \bibnamefont {Simmons}}, \bibinfo
  {author} {\bibfnamefont {M.}~\bibnamefont {Pepper}}, \bibinfo {author}
  {\bibfnamefont {A.~C.}\ \bibnamefont {Churchill}}, \ and\ \bibinfo {author}
  {\bibfnamefont {D.~A.}\ \bibnamefont {Ritchie}},\ }\href
  {http://stacks.iop.org/0953-8984/7/i=39/a=010} {\bibfield  {journal}
  {\bibinfo  {journal} {Journal of Physics: Condensed Matter}\ }\textbf
  {\bibinfo {volume} {7}},\ \bibinfo {pages} {7675} (\bibinfo {year}
  {1995}{\natexlab{b}})}\BibitemShut {NoStop}%
\bibitem [{\citenamefont {Rotter}\ \emph {et~al.}(1998)\citenamefont {Rotter},
  \citenamefont {Wixforth}, \citenamefont {Ruile}, \citenamefont {Bernklau},\
  and\ \citenamefont {Riechert}}]{Rotter_APL73_2128_98}%
  \BibitemOpen
  \bibfield  {author} {\bibinfo {author} {\bibfnamefont {M.}~\bibnamefont
  {Rotter}}, \bibinfo {author} {\bibfnamefont {A.}~\bibnamefont {Wixforth}},
  \bibinfo {author} {\bibfnamefont {W.}~\bibnamefont {Ruile}}, \bibinfo
  {author} {\bibfnamefont {D.}~\bibnamefont {Bernklau}}, \ and\ \bibinfo
  {author} {\bibfnamefont {H.}~\bibnamefont {Riechert}},\ }\href {\doibase
  10.1063/1.122400} {\bibfield  {journal} {\bibinfo  {journal} {Appl. Phys.
  Lett.}\ }\textbf {\bibinfo {volume} {73}},\ \bibinfo {pages} {2128} (\bibinfo
  {year} {1998})}\BibitemShut {NoStop}%
\bibitem [{\citenamefont {Thalmeier}, \citenamefont {D\'ora},\ and\
  \citenamefont {Ziegler}(2010)}]{Thalmeier_PRB81_41409_10}%
  \BibitemOpen
  \bibfield  {author} {\bibinfo {author} {\bibfnamefont {P.}~\bibnamefont
  {Thalmeier}}, \bibinfo {author} {\bibfnamefont {B.}~\bibnamefont {D\'ora}}, \
  and\ \bibinfo {author} {\bibfnamefont {K.}~\bibnamefont {Ziegler}},\ }\href
  {\doibase 10.1103/PhysRevB.81.041409} {\bibfield  {journal} {\bibinfo
  {journal} {Phys. Rev. B.}\ }\textbf {\bibinfo {volume} {81}},\ \bibinfo
  {pages} {041409(R)} (\bibinfo {year} {2010})}\BibitemShut {NoStop}%
\bibitem [{\citenamefont {Zhang}\ and\ \citenamefont
  {Xu}(2011)}]{Zhang_AA1_22146_11}%
  \BibitemOpen
  \bibfield  {author} {\bibinfo {author} {\bibfnamefont {S.~H.}\ \bibnamefont
  {Zhang}}\ and\ \bibinfo {author} {\bibfnamefont {W.}~\bibnamefont {Xu}},\
  }\href {\doibase http://dx.doi.org/10.1063/1.3608045} {\bibfield  {journal}
  {\bibinfo  {journal} {AIP Adv.}\ }\textbf {\bibinfo {volume} {1}},\ \bibinfo
  {eid} {022146} (\bibinfo {year} {2011})}\BibitemShut {NoStop}%
\bibitem [{\citenamefont {Insepov}\ \emph {et~al.}(2015)\citenamefont
  {Insepov}, \citenamefont {Emelin}, \citenamefont {Kononenko}, \citenamefont
  {Roshchupkin}, \citenamefont {Tnyshtykbayev},\ and\ \citenamefont
  {Baigarin}}]{Insepov_APL106_23505_15}%
  \BibitemOpen
  \bibfield  {author} {\bibinfo {author} {\bibfnamefont {Z.}~\bibnamefont
  {Insepov}}, \bibinfo {author} {\bibfnamefont {E.}~\bibnamefont {Emelin}},
  \bibinfo {author} {\bibfnamefont {O.}~\bibnamefont {Kononenko}}, \bibinfo
  {author} {\bibfnamefont {D.}~\bibnamefont {Roshchupkin}}, \bibinfo {author}
  {\bibfnamefont {K.~B.}\ \bibnamefont {Tnyshtykbayev}}, \ and\ \bibinfo
  {author} {\bibfnamefont {K.~A.}\ \bibnamefont {Baigarin}},\ }\href {\doibase
  10.1063/1.4906033} {\bibfield  {journal} {\bibinfo  {journal} {Appl. Phys.
  Lett.}\ }\textbf {\bibinfo {volume} {106}},\ \bibinfo {pages} {023505}
  (\bibinfo {year} {2015})}\BibitemShut {NoStop}%
\bibitem [{\citenamefont {Miseikis}\ \emph {et~al.}(2012)\citenamefont
  {Miseikis}, \citenamefont {Cunningham}, \citenamefont {Saeed}, \citenamefont
  {O'Rorke},\ and\ \citenamefont {Davies}}]{Miseikis_APL100_133105_12}%
  \BibitemOpen
  \bibfield  {author} {\bibinfo {author} {\bibfnamefont {V.}~\bibnamefont
  {Miseikis}}, \bibinfo {author} {\bibfnamefont {J.~E.}\ \bibnamefont
  {Cunningham}}, \bibinfo {author} {\bibfnamefont {K.}~\bibnamefont {Saeed}},
  \bibinfo {author} {\bibfnamefont {R.}~\bibnamefont {O'Rorke}}, \ and\
  \bibinfo {author} {\bibfnamefont {A.~G.}\ \bibnamefont {Davies}},\ }\href
  {\doibase 10.1063/1.3697403} {\bibfield  {journal} {\bibinfo  {journal}
  {Appl. Phys. Lett.}\ }\textbf {\bibinfo {volume} {100}},\ \bibinfo {pages}
  {133105} (\bibinfo {year} {2012})}\BibitemShut {NoStop}%
\bibitem [{\citenamefont {Bandhu}, \citenamefont {Lawton},\ and\ \citenamefont
  {Nash}(2013)}]{Bandhu_APL103_133101_13}%
  \BibitemOpen
  \bibfield  {author} {\bibinfo {author} {\bibfnamefont {L.}~\bibnamefont
  {Bandhu}}, \bibinfo {author} {\bibfnamefont {L.~M.}\ \bibnamefont {Lawton}},
  \ and\ \bibinfo {author} {\bibfnamefont {G.~R.}\ \bibnamefont {Nash}},\
  }\href {\doibase 10.1063/1.4822121} {\bibfield  {journal} {\bibinfo
  {journal} {Appl. Phys. Lett.}\ }\textbf {\bibinfo {volume} {103}},\ \bibinfo
  {pages} {133101} (\bibinfo {year} {2013})}\BibitemShut {NoStop}%
\bibitem [{\citenamefont {Bandhu}\ and\ \citenamefont
  {Nash}(2014)}]{Bandhu_APL105_263106_14}%
  \BibitemOpen
  \bibfield  {author} {\bibinfo {author} {\bibfnamefont {L.}~\bibnamefont
  {Bandhu}}\ and\ \bibinfo {author} {\bibfnamefont {G.~R.}\ \bibnamefont
  {Nash}},\ }\href {\doibase 10.1063/1.4905222} {\bibfield  {journal} {\bibinfo
   {journal} {Appl. Phys. Lett.}\ }\textbf {\bibinfo {volume} {105}},\ \bibinfo
  {pages} {263106} (\bibinfo {year} {2014})}\BibitemShut {NoStop}%
\bibitem [{\citenamefont {Roshchupkin}\ \emph {et~al.}(2015)\citenamefont
  {Roshchupkin}, \citenamefont {Ortega}, \citenamefont {Zizak}, \citenamefont
  {Plotitcyna}, \citenamefont {Matveev}, \citenamefont {Kononenko},
  \citenamefont {Emelin}, \citenamefont {Erko}, \citenamefont {Tynyshtykbayev},
  \citenamefont {Irzhak},\ and\ \citenamefont
  {Insepov}}]{Roshchupkin_JAP118_104901_15}%
  \BibitemOpen
  \bibfield  {author} {\bibinfo {author} {\bibfnamefont {D.}~\bibnamefont
  {Roshchupkin}}, \bibinfo {author} {\bibfnamefont {L.}~\bibnamefont {Ortega}},
  \bibinfo {author} {\bibfnamefont {I.}~\bibnamefont {Zizak}}, \bibinfo
  {author} {\bibfnamefont {O.}~\bibnamefont {Plotitcyna}}, \bibinfo {author}
  {\bibfnamefont {V.}~\bibnamefont {Matveev}}, \bibinfo {author} {\bibfnamefont
  {O.}~\bibnamefont {Kononenko}}, \bibinfo {author} {\bibfnamefont
  {E.}~\bibnamefont {Emelin}}, \bibinfo {author} {\bibfnamefont
  {A.}~\bibnamefont {Erko}}, \bibinfo {author} {\bibfnamefont {K.}~\bibnamefont
  {Tynyshtykbayev}}, \bibinfo {author} {\bibfnamefont {D.}~\bibnamefont
  {Irzhak}}, \ and\ \bibinfo {author} {\bibfnamefont {Z.}~\bibnamefont
  {Insepov}},\ }\href {\doibase 10.1063/1.4930050} {\bibfield  {journal}
  {\bibinfo  {journal} {J. Appl. Phys.}\ }\textbf {\bibinfo {volume} {118}},\
  \bibinfo {pages} {104901} (\bibinfo {year} {2015})}\BibitemShut {NoStop}%
\bibitem [{\citenamefont {Schiefele}\ \emph {et~al.}(2013)\citenamefont
  {Schiefele}, \citenamefont {Pedr\'os}, \citenamefont {Sols}, \citenamefont
  {Calle},\ and\ \citenamefont {Guinea}}]{Schiefele_PRL111_237405_13}%
  \BibitemOpen
  \bibfield  {author} {\bibinfo {author} {\bibfnamefont {J.}~\bibnamefont
  {Schiefele}}, \bibinfo {author} {\bibfnamefont {J.}~\bibnamefont {Pedr\'os}},
  \bibinfo {author} {\bibfnamefont {F.}~\bibnamefont {Sols}}, \bibinfo {author}
  {\bibfnamefont {F.}~\bibnamefont {Calle}}, \ and\ \bibinfo {author}
  {\bibfnamefont {F.}~\bibnamefont {Guinea}},\ }\href {\doibase
  10.1103/PhysRevLett.111.237405} {\bibfield  {journal} {\bibinfo  {journal}
  {Phys. Rev. Lett.}\ }\textbf {\bibinfo {volume} {111}},\ \bibinfo {pages}
  {237405} (\bibinfo {year} {2013})}\BibitemShut {NoStop}%
\bibitem [{\citenamefont {Dietel}\ and\ \citenamefont
  {Kleinert}(2012)}]{Dietel_PRB86_115450_12}%
  \BibitemOpen
  \bibfield  {author} {\bibinfo {author} {\bibfnamefont {J.}~\bibnamefont
  {Dietel}}\ and\ \bibinfo {author} {\bibfnamefont {H.}~\bibnamefont
  {Kleinert}},\ }\href {\doibase 10.1103/PhysRevB.86.115450} {\bibfield
  {journal} {\bibinfo  {journal} {Phys. Rev. B}\ }\textbf {\bibinfo {volume}
  {86}},\ \bibinfo {pages} {115450} (\bibinfo {year} {2012})}\BibitemShut
  {NoStop}%
\bibitem [{\citenamefont {Santos}\ \emph {et~al.}(2013)\citenamefont {Santos},
  \citenamefont {Schumann}, \citenamefont {Oliveira}, \citenamefont {Lopes},\
  and\ \citenamefont {Riechert}}]{Santos_APL102_221907_13}%
  \BibitemOpen
  \bibfield  {author} {\bibinfo {author} {\bibfnamefont {P.~V.}\ \bibnamefont
  {Santos}}, \bibinfo {author} {\bibfnamefont {T.}~\bibnamefont {Schumann}},
  \bibinfo {author} {\bibfnamefont {M.~H.}\ \bibnamefont {Oliveira}}, \bibinfo
  {author} {\bibfnamefont {J.~M.~J.}\ \bibnamefont {Lopes}}, \ and\ \bibinfo
  {author} {\bibfnamefont {H.}~\bibnamefont {Riechert}},\ }\href {\doibase
  10.1063/1.4809726} {\bibfield  {journal} {\bibinfo  {journal} {Appl. Phys.
  Lett.}\ }\textbf {\bibinfo {volume} {102}},\ \bibinfo {pages} {221907}
  (\bibinfo {year} {2013})}\BibitemShut {NoStop}%
\bibitem [{\citenamefont {Bandhu}\ and\ \citenamefont
  {Nash}(2016)}]{Bandhu_NR9_685_16}%
  \BibitemOpen
  \bibfield  {author} {\bibinfo {author} {\bibfnamefont {L.}~\bibnamefont
  {Bandhu}}\ and\ \bibinfo {author} {\bibfnamefont {G.~R.}\ \bibnamefont
  {Nash}},\ }\href {\doibase 10.1007/s12274-015-0947-z} {\bibfield  {journal}
  {\bibinfo  {journal} {Nano Research}\ }\textbf {\bibinfo {volume} {9}},\
  \bibinfo {pages} {685} (\bibinfo {year} {2016})}\BibitemShut {NoStop}%
\bibitem [{\citenamefont {Oliveira}\ \emph {et~al.}(2011)\citenamefont
  {Oliveira}, \citenamefont {Schumann}, \citenamefont {Ramsteiner},
  \citenamefont {Lopes},\ and\ \citenamefont
  {Riechert}}]{Oliveira_APL99_111901_11}%
  \BibitemOpen
  \bibfield  {author} {\bibinfo {author} {\bibfnamefont {M.~H.}\ \bibnamefont
  {Oliveira}}, \bibinfo {author} {\bibfnamefont {T.}~\bibnamefont {Schumann}},
  \bibinfo {author} {\bibfnamefont {M.}~\bibnamefont {Ramsteiner}}, \bibinfo
  {author} {\bibfnamefont {J.~M.~J.}\ \bibnamefont {Lopes}}, \ and\ \bibinfo
  {author} {\bibfnamefont {H.}~\bibnamefont {Riechert}},\ }\href {\doibase
  10.1063/1.3638058} {\bibfield  {journal} {\bibinfo  {journal} {Appl. Phys.
  Lett.}\ }\textbf {\bibinfo {volume} {99}},\ \bibinfo {pages} {111901}
  (\bibinfo {year} {2011})}\BibitemShut {NoStop}%
\bibitem [{\citenamefont {Hwang}\ \emph {et~al.}(2012)\citenamefont {Hwang},
  \citenamefont {Tahy}, \citenamefont {Nyakiti}, \citenamefont {Wheeler},
  \citenamefont {Myers-Ward}, \citenamefont {Eddy~Jr.}, \citenamefont
  {Gaskill}, \citenamefont {H.}, \citenamefont {Seabaugh},\ and\ \citenamefont
  {Jena}}]{Hwang_JVSTB30_3_12}%
  \BibitemOpen
  \bibfield  {author} {\bibinfo {author} {\bibfnamefont {W.~S.}\ \bibnamefont
  {Hwang}}, \bibinfo {author} {\bibfnamefont {K.}~\bibnamefont {Tahy}},
  \bibinfo {author} {\bibfnamefont {L.~O.}\ \bibnamefont {Nyakiti}}, \bibinfo
  {author} {\bibfnamefont {V.~D.}\ \bibnamefont {Wheeler}}, \bibinfo {author}
  {\bibfnamefont {R.~L.}\ \bibnamefont {Myers-Ward}}, \bibinfo {author}
  {\bibfnamefont {C.~R.}\ \bibnamefont {Eddy~Jr.}}, \bibinfo {author}
  {\bibfnamefont {D.~K.}\ \bibnamefont {Gaskill}}, \bibinfo {author}
  {\bibfnamefont {X.}~\bibnamefont {H.}}, \bibinfo {author} {\bibfnamefont
  {A.}~\bibnamefont {Seabaugh}}, \ and\ \bibinfo {author} {\bibfnamefont
  {D.}~\bibnamefont {Jena}},\ }\href {\doibase 10.1116/1.3693593} {\bibfield
  {journal} {\bibinfo  {journal} {J. Vac. Sci. Tech. B}\ }\textbf {\bibinfo
  {volume} {30}},\ \bibinfo {pages} {03D104} (\bibinfo {year}
  {2012})}\BibitemShut {NoStop}%
\bibitem [{APL()}]{APL_SM}%
  \BibitemOpen
  \href@noop {} {}\bibinfo {note} {See supplemental material at [URL will be
  inserted by AIP] for Raman measurements, SEM image of the IDT fingers,
  numerical simulation of spatial profiles of $s_{11}$ resonances, and
  time-resolved $s_{11}$ and $s_{12}$ of the 7.13~GHz mode.}\BibitemShut
  {Stop}%
\bibitem [{\citenamefont {Yuan}, \citenamefont {Yin},\ and\ \citenamefont
  {Ning}(2013)}]{Yuan_PSaT15_86_13}%
  \BibitemOpen
  \bibfield  {author} {\bibinfo {author} {\bibfnamefont {Q.}~\bibnamefont
  {Yuan}}, \bibinfo {author} {\bibfnamefont {G.}~\bibnamefont {Yin}}, \ and\
  \bibinfo {author} {\bibfnamefont {Z.}~\bibnamefont {Ning}},\ }\href {\doibase
  10.1088/1009-0630/15/1/14} {\bibfield  {journal} {\bibinfo  {journal} {Plasma
  Sci. Technol.}\ }\textbf {\bibinfo {volume} {15}},\ \bibinfo {pages} {86}
  (\bibinfo {year} {2013})}\BibitemShut {NoStop}%
\bibitem [{\citenamefont {Schumann}\ \emph {et~al.}(2012)\citenamefont
  {Schumann}, \citenamefont {Friedland}, \citenamefont {Oliveira~Jr.},
  \citenamefont {Tahraoui}, \citenamefont {Lopes},\ and\ \citenamefont
  {Riechert}}]{Schumann_PRB85_235402_12}%
  \BibitemOpen
  \bibfield  {author} {\bibinfo {author} {\bibfnamefont {T.}~\bibnamefont
  {Schumann}}, \bibinfo {author} {\bibfnamefont {K.-J.}\ \bibnamefont
  {Friedland}}, \bibinfo {author} {\bibfnamefont {M.~H.}\ \bibnamefont
  {Oliveira~Jr.}}, \bibinfo {author} {\bibfnamefont {A.}~\bibnamefont
  {Tahraoui}}, \bibinfo {author} {\bibfnamefont {J.~M.~J.}\ \bibnamefont
  {Lopes}}, \ and\ \bibinfo {author} {\bibfnamefont {H.}~\bibnamefont
  {Riechert}},\ }\href {\doibase 10.1103/PhysRevB.85.235402} {\bibfield
  {journal} {\bibinfo  {journal} {Phys. Rev. B}\ }\textbf {\bibinfo {volume}
  {85}},\ \bibinfo {pages} {235402} (\bibinfo {year} {2012})}\BibitemShut
  {NoStop}%
\bibitem [{\citenamefont {Nakahata}\ \emph {et~al.}(1994)\citenamefont
  {Nakahata}, \citenamefont {Higaki}, \citenamefont {Hachigo}, \citenamefont
  {Shikata}, \citenamefont {Fujimori}, \citenamefont {Takahashi}, \citenamefont
  {Kajihara},\ and\ \citenamefont {Yamamoto}}]{Nakahata_JJAP33_324_94}%
  \BibitemOpen
  \bibfield  {author} {\bibinfo {author} {\bibfnamefont {H.}~\bibnamefont
  {Nakahata}}, \bibinfo {author} {\bibfnamefont {K.}~\bibnamefont {Higaki}},
  \bibinfo {author} {\bibfnamefont {A.}~\bibnamefont {Hachigo}}, \bibinfo
  {author} {\bibfnamefont {S.}~\bibnamefont {Shikata}}, \bibinfo {author}
  {\bibfnamefont {N.}~\bibnamefont {Fujimori}}, \bibinfo {author}
  {\bibfnamefont {Y.}~\bibnamefont {Takahashi}}, \bibinfo {author}
  {\bibfnamefont {T.}~\bibnamefont {Kajihara}}, \ and\ \bibinfo {author}
  {\bibfnamefont {Y.}~\bibnamefont {Yamamoto}},\ }\href@noop {} {\bibfield
  {journal} {\bibinfo  {journal} {Jpn. J. Appl. Phys.}\ }\textbf {\bibinfo
  {volume} {33}},\ \bibinfo {pages} {324} (\bibinfo {year} {1994})}\BibitemShut
  {NoStop}%
\bibitem [{\citenamefont {Nakahata}\ \emph {et~al.}(1995)\citenamefont
  {Nakahata}, \citenamefont {Hachigo}, \citenamefont {Higaki},\ and\
  \citenamefont {Fujii}}]{Nakahata_ITUFFC42_362_95}%
  \BibitemOpen
  \bibfield  {author} {\bibinfo {author} {\bibfnamefont {H.}~\bibnamefont
  {Nakahata}}, \bibinfo {author} {\bibfnamefont {A.}~\bibnamefont {Hachigo}},
  \bibinfo {author} {\bibfnamefont {K.}~\bibnamefont {Higaki}}, \ and\ \bibinfo
  {author} {\bibfnamefont {S.}~\bibnamefont {Fujii}},\ }\href@noop {}
  {\bibfield  {journal} {\bibinfo  {journal} {IEEE Trans. Ultrason.
  Ferroelectrics, Freq. Cont.}\ }\textbf {\bibinfo {volume} {42}},\ \bibinfo
  {pages} {362} (\bibinfo {year} {1995})}\BibitemShut {NoStop}%
\bibitem [{\citenamefont {Naumenko}\ and\ \citenamefont
  {Didenko}(1999)}]{Naumenko_APL75_3029_99}%
  \BibitemOpen
  \bibfield  {author} {\bibinfo {author} {\bibfnamefont {N.~F.}\ \bibnamefont
  {Naumenko}}\ and\ \bibinfo {author} {\bibfnamefont {I.~S.}\ \bibnamefont
  {Didenko}},\ }\href@noop {} {\bibfield  {journal} {\bibinfo  {journal} {App.
  Phys. Lett.}\ }\textbf {\bibinfo {volume} {75}},\ \bibinfo {pages} {3029}
  (\bibinfo {year} {1999})}\BibitemShut {NoStop}%
\bibitem [{\citenamefont {Didenko}, \citenamefont {Hickernell},\ and\
  \citenamefont {Naumenko}()}]{Didenko_IToUFaFC47_179_}%
  \BibitemOpen
  \bibfield  {author} {\bibinfo {author} {\bibfnamefont {I.~S.}\ \bibnamefont
  {Didenko}}, \bibinfo {author} {\bibfnamefont {F.~S.}\ \bibnamefont
  {Hickernell}}, \ and\ \bibinfo {author} {\bibfnamefont {N.~F.}\ \bibnamefont
  {Naumenko}},\ }\href@noop {} {\bibfield  {journal} {\bibinfo  {journal} {IEEE
  Transactions on Ultrasonics, Ferroelectrics and Frequency Control}\ }\textbf
  {\bibinfo {volume} {47}},\ \bibinfo {pages} {179}}\BibitemShut {NoStop}%
\bibitem [{\citenamefont {Takagaki}\ \emph {et~al.}(2002)\citenamefont
  {Takagaki}, \citenamefont {Santos}, \citenamefont {Wiebicke}, \citenamefont
  {Brandt}, \citenamefont {Sch\"{o}nherr},\ and\ \citenamefont
  {Ploog}}]{PVS120}%
  \BibitemOpen
  \bibfield  {author} {\bibinfo {author} {\bibfnamefont {Y.}~\bibnamefont
  {Takagaki}}, \bibinfo {author} {\bibfnamefont {P.~V.}\ \bibnamefont
  {Santos}}, \bibinfo {author} {\bibfnamefont {E.}~\bibnamefont {Wiebicke}},
  \bibinfo {author} {\bibfnamefont {O.}~\bibnamefont {Brandt}}, \bibinfo
  {author} {\bibfnamefont {H.-P.}\ \bibnamefont {Sch\"{o}nherr}}, \ and\
  \bibinfo {author} {\bibfnamefont {K.~H.}\ \bibnamefont {Ploog}},\ }\href@noop
  {} {\bibfield  {journal} {\bibinfo  {journal} {Phys. Rev. B}\ }\textbf
  {\bibinfo {volume} {66}},\ \bibinfo {pages} {155439} (\bibinfo {year}
  {2002})}\BibitemShut {NoStop}%
\bibitem [{\citenamefont {Wu}\ and\ \citenamefont
  {Wang}(2004)}]{Wu_JAP96_5249_04}%
  \BibitemOpen
  \bibfield  {author} {\bibinfo {author} {\bibfnamefont {T.-T.}\ \bibnamefont
  {Wu}}\ and\ \bibinfo {author} {\bibfnamefont {W.-S.}\ \bibnamefont {Wang}},\
  }\href@noop {} {\bibfield  {journal} {\bibinfo  {journal} {J. Appl. Phys.}\
  }\textbf {\bibinfo {volume} {96}},\ \bibinfo {pages} {5249} (\bibinfo {year}
  {2004})}\BibitemShut {NoStop}%
\bibitem [{\citenamefont {Engan}(1969)}]{Engan_ITED16_1014_69}%
  \BibitemOpen
  \bibfield  {author} {\bibinfo {author} {\bibfnamefont {H.}~\bibnamefont
  {Engan}},\ }\href {\doibase 10.1109/T-ED.1969.16902} {\bibfield  {journal}
  {\bibinfo  {journal} {IEEE Trans. Electron Devices}\ }\textbf {\bibinfo
  {volume} {16}},\ \bibinfo {pages} {1014 } (\bibinfo {year}
  {1969})}\BibitemShut {NoStop}%
\bibitem [{\citenamefont {Campbell}(1998)}]{Campbell_98}%
  \BibitemOpen
  \bibfield  {author} {\bibinfo {author} {\bibfnamefont {C.~K.}\ \bibnamefont
  {Campbell}},\ }\href
  {http://books.google.de/books?id=RM6fvMhN4V0C&pg=PA46&lpg=PA46&dq=ST+quartz+surface+acoustic+euler&source=bl&ots=xrvJpmiD1j&sig=5KfberBbmUwYd_mhLkXxqG6CNUM&hl=de&ei=zCwdTeq7Ao6v8QPrg83EBQ&sa=X&oi=book_result&ct=result&resnum=10&ved=0CH4Q6AEwCQ#v=onepage&q&f=false}
  {\emph {\bibinfo {title} {Surface acoustic wave devices for mobile and
  wireless communications}}}\ (\bibinfo  {publisher} {Academic Press},\
  \bibinfo {year} {1998})\BibitemShut {NoStop}%
\bibitem [{\citenamefont {Vozmediano}, \citenamefont {Katsnelson},\ and\
  \citenamefont {Guinea}(2010)}]{Vozmediano_PR496_109_10}%
  \BibitemOpen
  \bibfield  {author} {\bibinfo {author} {\bibfnamefont {M.~A.~H.}\
  \bibnamefont {Vozmediano}}, \bibinfo {author} {\bibfnamefont {M.~I.}\
  \bibnamefont {Katsnelson}}, \ and\ \bibinfo {author} {\bibfnamefont
  {F.}~\bibnamefont {Guinea}},\ }\href {\doibase 10.1016/j.physrep.2010.07.003}
  {\bibfield  {journal} {\bibinfo  {journal} {Phys. Rep.}\ }\textbf {\bibinfo
  {volume} {496}},\ \bibinfo {pages} {109} (\bibinfo {year}
  {2010})}\BibitemShut {NoStop}%
\end{thebibliography}

%

\end{document}